# Cenozoic Uplift of south Western Australia as constrained by river profiles

N. Barnett-Moore, N. Flament, C. Heine, N. Butterworth, R. D. Müller

*Earthbyte Group, School of Geosciences, The University of Sydney, NSW 2006, Australia, Email: nicholas.barnett-moore@sydney.edu.au*

## Abstract

The relative tectonic quiescence of the Australian continent during the Cenozoic makes it an excellent natural laboratory to study recent large-scale variations in surface topography, and processes that influence changes in its elevation. Embedded within this topography is a fluvial network that is sensitive to variations in horizontal and vertical motions. The notion that a river acts as a 'tape recorder' for vertical perturbations suggests that changes in spatial and temporal characteristics of surface uplift can be deduced through the analysis of longitudinal river profiles. We analyse 20 longitudinal river profiles around the Australian continent. Concave upward profiles in northeast Australia indicate an absence of recent surface uplift. In contrast, the major knickzones within longitudinal profiles of rivers in southwest Australia suggest recent surface uplift. Given the lack of recent large-scale tectonic activity in that region, this uplift requires an explanation. Applying an inverse algorithm to river profiles of south Western Australia reveals that this surface uplift started in the Eocene and culminated in the mid-late Neogene. The surface uplift rates deduced from this river profile analysis generally agree with independent geological observations including preserved shallow-marine sediment outcrops across the Eucla Basin and south Western Australia. We show that the interplay between global sea level and long-wavelength dynamic topography associated with south Western

Australia's plate motion path over the remnants of an ancient Pacific slab is a plausible mechanism driving this surface uplift.

# 1. Introduction

The Australian continent displays remarkable intermediate ($10^2$ km) to long-wavelength ($10^3$ km) tectonic stability throughout the Cenozoic. Since its break-up from Antarctica along the Great Australian Bight and the opening of the Tasman Sea along the eastern margin in the Cretaceous (Veevers, 1984), Australia has been tectonically relatively quiescent, with vertical surface displacements of the northward-moving continent largely controlled by long-wavelength dynamic topography (Heine et al., 2010, Müller et al., 2000). Past Australian inundation patterns, deduced from preserved ancient shallow-water sediments (Langford et al., 1995), generally differ from global sea level trends and have commonly been attributed to the effects of mantle convection-induced dynamic topography (Czarnota et al., 2013, Matthews et al., 2011, Heine et al., 2010, DiCaprio et al., 2009, Sandiford, 2007, Gurnis et al., 1998, Veevers, 1984, Liu, 1979).

While subsidence over geological time scales is generally well preserved in the stratigraphic record, uplifting areas are subject to erosion and tend to have a poorer preservation potential (Flament et al., 2013, Olen et al., 2012). Here, tectonic geomorphology can be used to infer rates and patterns of surface uplift over geological timescales from information contained in the present-day fluvial network (Whipple and Tucker, 1999). In particular, longitudinal river profiles may indicate whether surface uplift with respect to sea level (England and Molnar, 1990) has affected a catchment area (Snyder et al., 2000). The analysis of longitudinal river profiles is generally applied to tectonically active regions where surface uplift rates can be estimated independently and compared with bedrock erosion rates (Schoenbohm et al., 2004, Snyder et al., 2000). Recently, Pritchard et al. (2009) and Roberts and White (2010) suggested that the present-day geometry of longitudinal river profiles contains time-dependent information pertaining to the evolution of landscape vertical motions over larger spatial and temporal scales (i.e., ~1–100+ Myr, 10–1000 km; Roberts et al., 2012) in tectonically quiescent regions. In this method, time-dependent surface uplift rates are estimated by parameterizing the elevation of a river profile as a function of its length (Pritchard et al., 2009). Indeed, surface uplift results in rapid changes in gradient near the river mouth that, over time, migrate upstream as knickpoints (Whipple and Tucker, 1999). Depending on retreat rate, knickpoints may be preserved in present-day longitudinal river profiles, providing information on past uplift events.

Here, we analyse 20 longitudinal river profiles across the Australian continent (Fig. 1). While profiles from northern Australia do not show evidence for anomalous vertical motions, the shape of river profiles in south Western Australia suggests that recent surface uplift on regional scale may have occurred. This was previously recognised by Cope (1975). The main proposed mechanisms for this regional-scale uplift in south Western Australia is long-wavelength dynamic topography and associated continent-wide tilting (Jakica et al., 2011, Quigley et al., 2010, Sandiford, 2007)

We apply the method of Pritchard et al. (2009) to south Western Australian rivers to constrain the timing of surface uplift in that region. We then discuss potential driving mechanisms of this uplift, integrating geological constraints on seismicity, marine deposition, tectonic activity and the northward motion of the Australian plate since the Eocene.

## 2. Methodology

We analyse 20 individual longitudinal river profiles grouped into five representative regions: south Western Australia, Pilbara, north Western Australia and Northern Territories, and north Queensland (Fig. 1). To ensure our study focuses on the effects of mantle-driven processes on landscape evolution, we have excluded the Southeast Highlands (Wellman, 1974, Wellman, 1979),

Flinders Ranges (Célérier et al., 2005) and Tasmania (Solomon et al., 1962) where this effect is considered eclipsed by shorter-wavelength tectonic processes. Tectonically induced vertical motions are generally much larger than that produced by mantle convective processes, making the latter difficult to identify (e.g. Flament et al., 2014).

## 2.1 Extraction, selection and geometry of longitudinal river profiles

### 2.1.1. Profile extraction

Each river profile was extracted from an SRTM 3 arc second DEM (Rabus et al., 2003). The DEM was segmented and reprojected into its respective UTM zones (50, 52 and 54) ensuring a consistent cell size of 90 m. A global assessment of the SRTM data indicates an absolute height error of 6.0 m, and relative height error of 4.7 m for the Australian continent (Rodriguez et al., 2006). We followed standard protocols to extract river profiles using the *Hydrology Tool* in ESRI ArcGis 10.0®. We first removed all anomalous spikes and troughs, ensuring a hydrologically sound DEM. Next; we established a drainage network using a standard flow-routing algorithm that determines the direction of flow via the steepest slope from each cell. This was then used to calculate the flow accumulation based on the cumulative weight of all cells flowing into each downslope cell. All profiles exceed Strahler stream order > 4 (Strahler, 1957) which defines stream size based on a hierarchy of tributaries. All rivers drain to the coastline, which is assumed to be the fixed reference level in this approach (Pritchard et al., 2009).

### 2.1.2. Profile selection

Rivers draining internally are excluded as they may experience changes in reference levels at their mouth, which is not located at sea level. Furthermore, rivers draining expansive inland regions may cross different swells and depressions, which would not be consistent with the assumption that drainage planforms do not vary and that uplift varies solely as a function of time (Pritchard et al., 2009). Such rivers were avoided because they would require more advanced modelling, such as a general 2D scheme that considers variations in uplift as a function of time and space (Czarnota et al., 2014, Roberts and White, 2010). We have also avoided, where possible, river profiles affected by dams (Kollmorgen et al., 2007) or rock uplift resulting in river captures and drainage reorganisations. For instance, the Swan/Avon and Moore rivers (profiles a-b, Fig. 1) experienced significant drainage reorganisation as the result of Eocene upwarping that produced a marginal north-south swell in south Western Australia (Beard, 1999, Beard, 2003). Finally, the reactivation of faults may perturb or control river profiles, resulting in localised rather than broad knickzones (> 100 km; Whittaker, 2012, Whittaker and Boulton, 2012). The Murchison River longitudinal profile (profile 7, Fig. 1) displays such a localised knickzone in close proximity to numerous Neogene reactivated faults (Fig. 1 - inset, yellow faults; Clark et al., 2012), suggesting local tectonics may have influenced the evolution of this river that we therefore exclude from our analysis. After careful analysis of published data (Clark et al., 2012), we assume that fault reactivation has not controlled the evolution of other rivers. Final river profiles were compared to surface hydrology maps (BoM, 2013) to ensure their validity.

### 2.1.3. Profile geometry

The quality-controlled profiles were analysed to determine important morphological features that allude to the influences of landscape vertical motions. Using the geomorphology software *Geomorph Tools* (Whipple et al., 2007), we quantitatively identified all minor and major knickzones through a combination of longitudinal plots (Fig. 1) and linear regression of logarithmic slope versus drainage area plots (Snyder et al., 2003). A major knickzone is defined as a change in profile slope downstream from the inflexion point (Wobus et al., 2006, Whipple and Tucker, 1999). Out of the numerous classifications of knickzones and knickpoints (Foster and Kelsey, 2012, Goldrick and Bishop, 1995), we focused on first order knickzones, neglecting knickpoints that may have formed via spatial contrasts in lithology (VanLaningham et al., 2006) or fault-related vertical motions that introduce higher amplitude knickpoint or knickzone geometries within longitudinal river profiles (Whittaker, 2012).

## 2.2. Parameterization of uplift history from longitudinal river profiles: Governing equations

The wavelength of uplift associated with dynamic topography is expected to be longer than the total length of any of the selected longitudinal river profiles. We therefore implemented a simple 1D inverse algorithm to determine uplift rates from the shape of longitudinal river profiles (Pritchard et al., 2009). In this approach, it is assumed that the elevation along a river profile is controlled by uplift, and moderated by advective erosion (Pritchard et al., 2009). It is known that the evolution of elevation along a longitudinal river profile, over time, can be written as

$$\frac{\partial z}{\partial t} = U(t) - v_o x^m (-\frac{\partial z}{\partial x})^n + k(x)\frac{d^2 z}{\partial x^2} \qquad (1)$$

where $U(t)$ is the uplift rate, $v_o$ is the reference knickpoint retreat velocity for $m = 0$ and $n = 1$, $m$ and $n$ are dimensionless parameters representing the distance and slope exponent, $k$ is the diffusivity, and $x^m$ the discharge, which increases downstream (Pritchard et al., 2009). Alternatively, $A^m$ is sometimes used as a proxy for discharge, where $A$ is the upstream drainage area at any position $x$ *(e.g.* Whipple and Tucker, 1999). A previous comparison between the two proxies of discharge showed $x^m$ to be adequate for Africa, with the important advantage of easier implementation (Roberts and White, 2010).

Based on Eq. (1), neglecting the diffusive term, and given profile information $z(x)$ at $t = 0$, it can be shown that (Pritchard et al., 2009)

$$U(\tau) = v_o x^m (-\frac{dz}{dx})^n \qquad (2)$$

where $\tau$ is the characteristic time period of uplift that can be expressed as

$$\tau = \frac{L^{1-m}}{(n-m)v_0}\left[\frac{1-\left(\frac{x}{L}\right)^{1-\frac{m}{n}}}{\left(\frac{x}{L}\right)^{\frac{m(n-1)}{n}}}\right](-\frac{dz}{dx})^{1-n} \qquad (3)$$

where $L$ represents the total length of a longitudinal river profile.

Eqs. (2) and (3) allow for the calculation of an uplift history from longitudinal river profiles. The resolution of the SRTM DEM (3 arc seconds) gives a characteristic time step length of ~45 kyr and determines the discrete spacing of calculated uplift rate, U ( $\tau$ ). Calculated uplift rate histories were smoothed using a Gaussian filter over an 8-Myr time window. Cumulative uplift was calculated by integrating the uplift rate history (Eq. 2) of each river over geological time.

## 2.3. Parameter selection, sensitivity study and model limitations

Here we justify our parameter selection and discuss the limitations associated with model assumptions, including the effects of underlying lithology and climate.

### 2.3.1. Parameter selection

Given the known difficulties accurately constraining the slope exponent $n$ in Eq. 1-3 (Whipple and Tucker, 1999), we followed Roberts and White (2010) in assuming $n$ = 1. This simplifies the problem as the advective term becomes linear, and calculated uplift rate histories then scale with $v_o$ (Eq. 2). In this simplified approach the advective velocity represents a decrease in the knickzone velocity as it propagates further upstream. Previous efforts linking a decrease in upstream drainage area, as a river approaches its source, with a decrease in knickzone velocities (Crosby and Whipple, 2006), suggest using $n$ = 1 is reasonable. Increasing $n$ to 1.05 would push $U(\tau)$ back in time by ~2-3 Myr. If $n$ > 1.05, multiple values of $U(\tau)$ may occur at a given time and Eqs. (2) and (3) would no longer apply (Pritchard et al., 2009).

Uplift rates and characteristic uplift period depend on the distance exponent $m$ and knickzone retreat rate $v_o$ (Eqs. 2 and 3). We calibrated $m$ and $v_o$ using the deposition age (~33.5-36.5 Ma) and present-day elevation range (100-250 m) of the Upper Eocene shallow-marine deposits of the Pallinup and Princess Royal Spongolite Formation (Gammon et al., 2000). This geological constraint implies that the cumulative uplift for the Young, Phillips, and Gardiner rivers should be between 100 and 250 m since ~36.5-33.5 Ma (white box in Fig. 5A). Systematically varying $m$ and $v_o$ within the range of published estimates, we selected $m = 0.5$ and $v_o = 5$ m$^{(1-m)}$ Myr$^{-1}$ (Fig. 5A).

A variation in the position of the coastline affects the total length ($L$) of a river profile (Pritchard et al., 2009). Coastline fluctuations of ±45 km change the calculated uplift rate history by ±1.5 Myr. Paleogeographic data indicates that the coastline position changed by ~50–70 km around south Western Australia since the Eocene (Langford et al., 1995), suggesting a temporal error of ~±2 Myr. Similarly, a linear vertical relative height error of < 10 m (Rodriguez et al., 2006) in the SRTM 3 arc second DEM results in an approximate uplift error of less than 1 m Myr$^{-1}$ in calculated uplift rate histories.

## 2.3.2. Sensitivity study

A sensitivity analysis of the distance exponent $m$ and knickzone retreat rate $v_o$ (Eqs. 2 and 3) was conducted to determine variations in calculated uplift rate histories, given different combinations of parameters. These were allowed to vary within the range of published estimates, for the example of the Gardiner River, to investigate model sensitivity to parameter selection (Fig. 2). The calculated uplift rate increases with $v_o$ and $m$, as expected (Fig. 2A), whereas the characteristic uplift period decreases with increasing $v_o$ and $m$ (Fig. 2B). Both the uplift rate and the characteristic uplift period vary by one order of magnitude for combinations of $m$ and $v_o$ (Fig. 2). Such variability is comparable to the variability in estimating exhumation rates, defined as the difference between surface and rock uplift (England and Molnar, 1990), derived from thermochronology (Kohn et al., 2002). This significant variability across reasonable combinations of $m$ and $v_o$ confirms that these parameters should be carefully selected based on independent geological data or previous studies (Roberts and White, 2010).

## 2.3.3. Potential limitations: Lithology and Climate

The absence of correlation between major knickzone locations and changes in underlying lithology (Fig. 3) — with the possible exception of the Phillips River (Fig. 3B) — suggests that these knickzones did not form because of spatial contrasts in erodibility (Miller et al., 2012) and that lithology has little effect on the shape of these longitudinal river profiles (Fig. 3). The cumulative uplift predicted for the Phillips River exceeds the maximum present-day elevation of the Eocene Pallinup and Princess Royal Spongolite Formations (Fig. 5A), further suggesting that the underlying lithology may have influenced the evolution of this particular river profile.

Knickpoint retreat rates also depend on climate, which is implicitly assumed to be constant in the method described above. The long-standing debate concerning the relative contributions of tectonics and climate on river profiles illustrates the difficulty associated with deconvolving these complex and non-linear interactions (Raymo and Ruddiman, 1992, Hren et al., 2007). Quantifying the climatic controls on topography still eludes scientists and current efforts mainly focus in regions of high precipitation rates (Stark et al., 2010), or consisting of mountainous terrain with steep drainage gradients (Ferrier et al., 2013, Burbank et al., 2003). D'Arcy and Whittaker (2013) suggested that precipitation rates control the steepening of river channels in response to tectonic uplift. Steep river channels downstream of knickzones only occur for the Phillips (Fig. 3B) and Collie (Fig. 3F) rivers, suggesting that climate does not control the channel steepness of south Western Australian rivers (D'Arcy and Whittaker, 2013). We also note the very similar profile geometry between the Swan, Moore, and Murchison rivers (Fig. 1), which were excluded from this analysis, and other river profiles in south Western Australia. However, the geometry of the Murchison and Gascoyne river profiles (Fig. 1) are very different. Given the proximity of these two rivers (~ 300 km) and the absence of significant relief between them, we argue that sharp contrasts in climate are unlikely to explain the difference in the shape of these rivers. An increase in mean annual precipitation in a tectonically stable region may lead to steeply concave river profiles (Zaprowski et al., 2005). However, northwest Australian river profiles are smooth, nearly flat (e.g. Gascoyne, Ashburton, DeGray; Fig. 1), when visually compared with northeast Australian river profiles (e.g., Flinders, Staaten, Leichhardt; Fig. 1), again suggesting a lack of climatic control across the tectonically stable region. Finally, Wobus et al. (2010) argued that climatic shifts could cause knickpoints to incise downstream and generate decreasing gradients downstream of the knickzone, which is not observed in the longitudinal river profiles of south Western Australia.

The Oligocene–Neogene marked a transition from the Cretaceous–Eocene paleoclimate of Australia (Quilty, 1994) towards its present-day climate. Along the southern margin, the Cretaceous–Eocene represented a period of climatic fluctuations influenced by Australia's proximity to Antarctica (Quilty, 1994). Changes to ocean currents as a result of Australia's migration away from Antarctica (Whittaker et al., 2007), coupled with the opening of the Drake passage between South America and Antarctica (Lawver and Gahagan, 1998) resulted in the establishment of the Circum-Antarctic current (Lawver and Gahagan, 2003) that increased coastal precipitation and runoff (Quilty, 1994). The present-day precipitation rates of 500-600 mm yr[-1] (BoM, 2013) in south Western Australia are one order of magnitude lower than required for climate to be the primary control on the evolution of longitudinal river profiles ($\approx$ 50 mm/day; Stark et al., 2010). Recent modelling determined that monsoon precipitations rates in north and western Australia were even lower during the Early to Middle Miocene (~200 mm yr[-1]) than at present (Herold et al., 2011). A change in climatic regimes thus appears unlikely to have controlled the evolution of south Western Australian rivers.

# 3. Results

## 3.1. Geomorphologic characteristics of Australian longitudinal river profiles

Concave longitudinal river profiles are usually interpreted to be in steady state, unaffected by external forces such as recent changes in erosion, uplift, or in climate (Whipple and Tucker, 1999). Examples from our study are the rivers in the north Queensland region, which appear to have developed in the recent absence of significant changes in these external controls (Fig. 1). In contrast, longitudinal river profiles displaying convex geometries, suggest disequilibrium as a result of the influence of one or more external forces (Whipple and Tucker, 1999). This is the case for river profiles from south Western Australia (Fig. 1) that all show convex shapes and knickzones at varying distances upstream from the river mouth. Interestingly, prominent south Western Australian knickzones all occur at a similar altitude (~200 ±20 m), suggesting that a uniform surface uplift event may have affected those longitudinal river profiles contemporaneously (Berlin and Anderson, 2007). Niemann et al. (2001) showed that in the absence of transport-limited erosion, and of spatial heterogeneities in uplift rate or erodibility, the knickzone retreat velocity should be regionally consistent such that knickzones resulting from a particular vertical perturbation should be found at the same elevation within a basin. Therefore, knickzones occurring at the same elevation across an area suggest spatially uniform uplift and a lack of transport-limited erosion (Niemann et al., 2001).

Prominent knickzones can be identified in river profiles occurring as far north as the Murchison River (Profile 7, Fig. 1), then progressively change to smooth concave upward geometries in the Pilbara region (Fig. 1). This concave upward profile geometry becomes more pronounced towards the Kimberley region and North Queensland. Here, longitudinal river profiles consistently have distinct concave upward profiles, suggesting they may be in equilibrium (Whipple and Tucker, 1999). Several studies have proposed that northern Australia is currently undergoing a dynamic drawdown, caused by Australia overriding slabs subducting under South-East Asia and Melanesia (Heine et al., 2010, DiCaprio et al., 2010, Müller et al., 2000). However, such dynamic subsidence cannot be quantified by the analysis of longitudinal river profiles that may only constrain uplift (Pritchard et al., 2009). Finally, the measured concavities ($\theta$; Whipple et al., 2007) of relict streams, i.e. the sections of a river profile upstream of a knickzone or knickpoint, unaffected by its migration (Schoenbohm et al., 2004) in south

Western Australia ($\theta$ = 0.38 ±0.3119) match the full-profile concavities measured in north Queensland ($\theta$ = 0.36 ±0.1225), suggesting that south Western Australia rivers might have once shared a similar profile geometry to the present-day river profiles of north Queensland.

## 3.2. Predicted uplift histories deduced from longitudinal river profiles

### 3.2.1. Predicted uplift rate history

Based on the analysis of considered river profiles (previous section and Fig. 1), we limit the application of the inverse algorithm of Pritchard et al. (2009) to south Western Australian rivers that consistently display convex profiles with pronounced knickzones. We excluded the Swan/Avon and Moore rivers (profiles a-b, Fig. 1) from this analysis because a Late Eocene drainage reversal has been reported (Beard, 1999) for these rivers.

Calculated uplift rates are displayed from 100 Ma to present day (Fig. 4) because there is no significant change in cumulative uplift prior this time (Fig. 5A).

The Phillips, Collie, and Franklin Rivers show uplift rates ranging between 0 and 3 m Myr$^{-1}$ until ~40 Ma. The Young, Gardiner, and Blackwood Rivers show minor undulations in uplift rates until ~45 Ma, not exceeding peak rates of 3 to 5 m Myr$^{-1}$.

The predicted uplift rate for the Young River (Fig. 4A) first increases from ~45 Ma, recording a small peak at ~40 Ma (~6 m Myr$^{-1}$). Uplift rates then plateau off at ~5 m Myr$^{-1}$, and increase again from 15 Ma onwards, reaching a maximum of 11 m Myr$^{-1}$ at ~10 Ma. Uplift rates decline sharply to present day (~3 m Myr$^{-1}$).

The Phillips River (Fig. 4B) records a sharp increase from 40 Ma until it reaches a peak rate of 21 m Myr$^{-1}$ at ~17 Ma, which is the maximum predicted uplift rate for any of the selected river profiles. As with the Young River, uplift rates decline immediately after 17 Ma to present day values of ~6 m Myr$^{-1}$.

The Gardiner River (Fig. 4C) indicates an increase of uplift rates starting at ~45 Ma, reaching peak rates at ~32 Ma (6 m Myr$^{-1}$) and at ~12 Ma (9 m Myr$^{-1}$) with a period of ~10 Myr of lower uplift rates separating the two maxima. Rates subsequently decline to present day (~3 m Myr$^{-1}$).

The predicted Franklin River uplift history (Fig. 4D) shows a slow increase from ~40 Ma, followed by a sharp increase from 22 Ma onwards with peak rates reached at ~18 Ma (10 m Myr$^{-1}$) and ~6 Ma (11 m Myr$^{-1}$). Given their temporal proximity and the negligible decline in uplift rates between them, these peaks represent a singular uplift event commencing at ~22 Ma. Rates then decline to present day (~3 m Myr$^{-1}$).

Along the Blackwood river (Fig. 4E), the model shows an increase in uplift rates from ~47 Ma onwards, peaking at ~45 Ma (~7 m Myr$^{-1}$). This rate remains relatively constant until ~32 Ma where it decreases to relatively low rates of ~3 m Myr$^{-1}$ until present day.

The predicted uplift history of the Collie River (Fig. 4F) is similar to that of the Phillips River with a sharp increase at ~25 Ma reaching ~20 m Myr$^{-1}$ around 16 Ma. The rate then declines sharply to ~1 m Myr$^{-1}$ at 5 Ma and remains constant to present day.

In summary, predicted uplift histories for the six river profiles in south Western Australia show a significant increase in uplift rates from ~45-40 Ma onwards, and a decrease in rate in the last 5 Ma (Fig. 4). There are variations between rivers in the timing and magnitude of recorded maximum uplift events between 40 and 5 Ma. The Young and Gardiner (12 Ma peak) rivers record maximum uplift rates of 10 m Myrs$^{-1}$ at around 10 Ma. The Phillips, and Collie rivers display the largest predicted spike in uplift rates of all profiles at ~16 Ma. The maximum uplift rates of the Blackwood River are consistent with the early rate increases of the Young, Phillips, and Gardiner Rivers at ~ 40 Ma.

Together, these results suggest that south Western Australia Rivers recorded an uplift event commencing in the mid-late Eocene with maximum uplift rates recorded in the Mid Neogene.

3.2.2. Predicted cumulative uplift

The cumulative uplift history is the total uplift over the predicted recording time of a longitudinal river profile (Fig. 5A) that depends on river length. We compute the evolution of total uplift normalised to present-day for all rivers (Fig. 5A). This reveals that south Western Australian rivers have recorded ~400 m of uplift over their history, with half of this uplift occurring since the mid-late Eocene (~200 m; 45–40 Ma). The initially low predicted uplift rates result in an increase of cumulative uplift of ~< 100 m for the first ~160 Myr until ~40 Ma. An exception is the Blackwood River (Fig. 5A, yellow line) that shows a slight acceleration in cumulative uplift from ~80 Ma. All other profiles show a sharp acceleration in cumulative uplift starting between 50 and 40 Ma, as expected from the predicted increase in uplift rates at that time (Fig. 5A). Individual cumulative histories differ in their gradient from this time onward, and for some rivers a second acceleration in cumulative uplift is recorded (e.g. Franklin River at ~20 Ma, Fig. 6a, purple). A later deceleration in cumulative uplift, is observed across all river profiles at ~10-5 Ma.

# 4. Discussion

## 4.1. Predicted uplift in the context of the geological record of southern Australia

The model cumulative uplift was calibrated to match the present-day elevation of the Upper Eocene shallow-marine deposits of the Pallinup and Princess Royal Spongolite Formation (Gammon et al., 2000). In addition, the preservation of a Late Eocene paleoshoreline at an elevation of ~300 m across the Eucla Basin, east of south Western Australia (Fig. 6A - black outline; Sandiford, 2007), suggests Late Eocene surface uplift further along the southern margin of Australia. This paleoshoreline includes inundated valleys, offshore barrier systems, and marginal lagoons all dating back to the mid-late Eocene (~41–39 Ma), indicating ~300 m of surface uplift across the Eucla Basin (~41-39 Ma;

Sandiford, 2007). Cope (1975) proposed parts of south Western Australia had undergone a primary phase of surface uplift during the Oligocene (~175 m) and a secondary event at the Miocene-Pliocene boundary, by measuring present-day elevations of late Eocene-Oligocene shallow-marine sediments. Previous to this, Lowry (1970), also measured present-day elevations of Eocene-Oligocene and Mid-Miocene shallow-marine sediments, suggesting two main phases of epeirogenic surface uplift in the region in the Oligocene and mid-Miocene. The preservation of mid-Miocene marine limestone within the Eucla Basin (Nullarbor Limestone; Li et al., 2003) at elevations of 180 – 280 m indicates subsequent surface uplift of that region during the Neogene (Fig. 6A; Sandiford, 2007). Further west, the preservation of the mid-Miocene Plumbridge limestone, a facies equivalent of the Nullarbor limestone (Lowry, 1970), crops out at elevations of ~50–100 m, indicating a similar surface uplift event during the Neogene influenced south Western Australia (Fig. 6).

The progressively increasing uplift rates from ~45–40 Ma derived from longitudinal profiles of south Western Australian rivers is generally compatible with present-day elevated outcrops of shallow-marine sediments across the southern margin of the Australian continent. Predicted cumulative uplift is in agreement with independent estimations of surface uplift from elevated outcrops of Oligocene (~175 m ± 15 m, 34-23 Ma; Cope, 1975) and mid-Miocene (~75 ± 25 m, 17-13 Ma; Lowry, 1970) shallow-water sediments (Fig. 5A; cyan). In this instance, the error bars shown for the Oligocene and mid-Miocene locations (Fig. 5A) do not account for any error in measurement, rather the variations in elevation of the respective outcrops. Late Eocene-Oligocene strata were deposited between ~34 and 23 Ma, while the mid-Miocene limestone was deposited at ~15 ± 2 Ma (Sandiford, 2007). Cumulative uplift histories are also compatible with the preservation of shallow-marine sediments along the onshore margin of the Bremer Basin and preserved paleoshorelines in the Eucla Basin. The maximum predicted uplift phase occurs in the mid- to late- Neogene, consistent with the preservation of mid-Neogene (~15 Ma) marine limestone to the East across the Eucla Basin, and in parts of south Western Australia (Fig. 6A). The protracted initial phase of uplift predicted by the inversion of river profiles cannot be validated in the absence of geological constraints on the pre-Eocene history of regional surface uplift.

The thin green curves in Fig. 4 represent smoothed long-term exhumation rates since 100 Ma for south Western Australia, derived from the analysis and modelling of apatite fission track on samples from ~120 locations in south Western Australia, averaged over 300 Myr (Kohn et al., 2002). Two distinct peaks in exhumation rates occur at ~35 Ma (~12 m $Myr^{-1}$) and ~25 Ma (~8 m $Myr^{-1}$). The first peak in exhumation rates (~35 Ma) occurs shortly after the initial increase in uplift rates recorded in the longitudinal river profiles (~45–40 Ma), and there is a significant difference in peak magnitudes (river surface uplift rates ~1–5 m $Myr^{-1}$ and exhumation rates ~12 m $Myr^{-1}$). The maximum uplift rates recorded in longitudinal river profiles occur later in geological time (Early – Mid Neogene) than the maximum exhumation rates (Late Eocene – Oligocene). Although poorly constrained given the assumptions underlying thermochronology and river studies, this time lag suggests that fluvial erosion may be an important mechanism of exhumation.

## 4.2. Mechanisms driving the uplift of south Western Australia

The evolution of topography results from multiple processes operating at different spatial and temporal scales, from low-amplitude (<2 km), long-wavelength (>700 km) dynamic topography (*e.g.* Flament et al, 2013), to large-amplitude (<10 km), short-wavelength (< 500 km) tectonic topography (Molnar and England, 1990), moderated by erosion and changes in eustatic sea level. Our analysis of south Western Australian longitudinal river profiles reveals ~ 200 m of surface uplift since ~45-40 Ma over a spatial scale of at least 500 km extending from the Young River to the Collie River (Figs. 1 and 3). We subsequently discuss processes operating at spatio-temporal wavelengths that could explain this surface uplift.

### 4.2.1. Long-wavelength dynamic topography

Continental-scale dynamic topography occurs at wavelengths (> 1000 km) compatible with the constrained surface uplift of south Western Australia (> 500 km) and has been previously recognized as a mechanism of surface uplift for this region (Czarnota et al., 2013, Quigley et al., 2010). Based on the pronounced latitudinal asymmetry of Neogene stratigraphy (Fig. 6B) and of the present-day Australian continental shelf, Sandiford (2007) estimated ~250–300 m of continental–scale dynamic north-down tilt since the mid-Miocene, at a rate of ~15–20 m Myr$^{-1}$. The preservation of a relict, broad-scale (~1000 km), southern tilt in the drainage basins of the Western Plateau (Beard 1998, 2000) further points to long-wavelength dynamic topography. It suggests that prior to Late Eocene surface uplift, the fluvial networks draining south Western Australia were influenced by a depression along the southern margin. This is further supported by observations detailing a late Paleogene episode of subsidence in the region, resulting in the deposition of the Werrilup Formation, followed by Oligocene surface uplift (Gammon et al., 2000, Cope, 1975, Cockbain, 1968).

To quantify the effect of mantle convection-induced dynamic topography on south Western Australian rivers, we analyse a geodynamic model backward advecting seismic tomography (Heine et al. 2010), shown since 40 Ma in a fixed Australian reference frame (Fig. 7). The evolution of Australian dynamic topography is dominated by the migration of the plate over two dynamic topography lows associated with sinking slabs. The southern margin of Australia was first drawn down (from ~40 Ma) by the large dynamic topography low associated with the sinking of the slab related to the subduction zone that separated Eastern Gondwanaland from the paleo-Pacific Ocean (Gurnis et al., 1998), then uplifted as the continent migrated northward (Fig. 7F – 7B). From ~ 25 Ma, the northward motion of Australia towards the dynamic topography low associated with Melanesian slabs (Müller et al., 2000) tilted the continent to the north (Fig. 7C – 7A).

The rate of change of dynamic topography predicted by this model (Fig. 4 - thin grey rate; Heine et al., 2010) shows a phase of dynamic surface uplift between ~30 and 15 Ma that is broadly compatible with a period of fast uplift rates recorded by the Phillips (Fig. 4B), Franklin (Fig. 4D) and Collie (Fig. 4F) rivers. However, uplift rates from all longitudinal river profiles initiate between ∼5–10 Myr earlier than this dynamic surface uplift (Fig. 4), and the change to dynamic subsidence from ∼15 Ma is at odds with the maximum uplift rates recorded by the Young (∼10 Ma), Phillips (∼18 Ma), Gardiner (∼12 Ma), Franklin (∼7 Ma and ∼20 Ma), and Collie (∼15 Ma) rivers. Parameters $m$ and $v_0$ could be changed in the river inversion algorithm (Eqs. 2-3) to fit the uplift history predicted by the geodynamic model, but the revised cumulative uplift would no longer be compatible with geological constraints (section 2.3.1 and Fig. 5A). This suggests that long-wavelength dynamic topography cannot explain the evolution of south Western Australian rivers if global sea level is assumed to be constant.

4.2.2. Interplays between eustasy and dynamic topography

Long-wavelength dynamic topography has been proposed to offset, by a third, the global sea level fall imposed by changes in the volume of ocean basins over the last 100 Ma (Spasojevic and Gurnis, 2012). However, the effect of dynamic topography on relative sea level varies from region to region (Spasojevic and Gurnis, 2012). We next consider fluctuations in eustatic sea level, using the curve of Haq and Al-Qahtani (2005; Fig. 7G), filtered for long-wavelengths (as it appeared in Muller et al., 2008). Interestingly, global sea level evolution is mostly synchronous with that of dynamic topography for south Western Australia (Fig. 7G). In geomorphology, base level is defined as the elevation below which a stream cannot incise (Leopold and Bull, 1979). However, here we calculate base level as the difference between dynamic topography and sea level. The evolution of base level is consistent with that of cumulative uplift, since uplift between ∼35 and 15 Ma is reflected by a regression (decrease in base level elevation, Fig. 5A). In addition, transgressions (increases in base level elevation) between ∼65 and 55 Ma and since ∼5 Ma (Fig. 4) correspond to constant or deceleration in cumulative uplift which confirms that the method used herein can only constrain uplift (Pritchard et al., 2009). While the trends in cumulative uplift and base level are consistent, amplitudes of changes in base level are about four times smaller than that of cumulative uplift. The change in base level since the Eocene (~90 m) is only half the present-day elevation of knickzones (~200 m), suggesting that the amplitude of dynamic topography, generally poorly constrained (e.g. Flament et al., 2013), might be underestimated in the geodynamic model used herein.

Rates of base level change (brown curves in Fig. 4) better agree with uplift rates constrained by river profiles than rates of dynamic topography change (grey curves in Fig. 4). Indeed, regressions (positive rates of base level change) at ~55 Ma, ~45 Ma, ~30 Ma, and between ~20 and 5 Ma are in good agreement with peaks in uplift rate at ~30 Ma and ~12 Ma (Gardiner River, Fig. 4C), ~45 Ma and ~30 Ma (Blackwood River, Fig. 4D) and consistent with maximum uplift rates occurring between ~20 and 5 Ma for all rivers except the Blackwood River. Transgressions (negative rates of base level change) at ~60 Ma, 40 Ma, 25 Ma and since ~5 Ma are consistent with decreasing or constant uplift rates at these times. As an example, uplift rates for the Phillips River plateau off between ~25 and 20 Ma, reflecting the transgression during this period.

The above analysis suggests that interplays between long-wavelength dynamic topography and eustasy are the primary mechanism controlling the evolution of south Western Australian rivers. Nevertheless, this comparison is not perfect and we discuss potential secondary mechanisms below.

### 4.2.3. Possible secondary mechanisms

Small discrepancies between the evolution of base level and uplift histories deduced from south Western Australian rivers, such as short-wavelength oscillations since 20 Ma in uplift rates predicted for the Franklin river, suggest that secondary mechanisms may be at play.

Changes in climate, although not controlling the evolution of south Western Australian rivers (see section 2.3.2), may play a secondary role as suggested by steep gradients downstream of knickzones for the Phillips and Collie rivers. Changes in climate could be included in more sophisticated models of river profiles, for instance by making knickpoint retreat rates time-dependent. This was not attempted here to keep the model simple, and not add a further uncertain parameter.

Changes in intraplate stresses across the Australian plate (Müller et al., 2012) could induce surface uplift over short (~1-5 Myr) time periods. Indeed, Miocene fault inversion and north-south anticlinal folding, attributed the collision of the Indo-Australian and Eurasian plates (Borissova et al., 2010, Iasky, 2003, Harris, 1994), occurs offshore Western Australia, but only north of ~28ºS (Kempton et al., 2011, Iasky, 2003). Onshore, the east-west orientation of the maximum horizontal stresses in the Yilgarn Craton should result in preferential reactivation of north–south striking structures, which is at odds with the Neogene seismic quiescence of the Darling Fault (Quigley et al., 2010, Clark and Leonard, 2003), a Cambrian tectonic feature bounding the eastern margin of the Perth Basin (Fig. 1; Veevers and Morgan, 2000). However, recent seismicity (1888-2000 earthquakes of magnitude > 5.5, Fig. 6A; Dent, 2008), indicates that some fault reactivation has occurred in south Western Australia. The distribution of this seismicity (Fig. 6A) suggests that lithospheric processes may influence Western Australian topography.

Edge-driven convection may result in vertical motions of time-scales of ~10 Myr and up to ~200 km outboard of a lithospheric step (King and Anderson, 1998), and this effect may be enhanced for a fast-moving plate (Farrington et al., 2010). This process may have occurred in south Western Australia given the fast motion of Australia since ~40 Ma (Whittaker et al., 2007), the steep gradient of lithospheric thickness between the Yilgarn Craton and the Perth Basin (Kennett and Salmon, 2012), and recent magmatic activity along the Western Australian margin (Gorter, 2009). Three-dimensional box models suggest that edge-driven convection at the trailing edge of a fast moving may result in vertical motions > 660 km downstream from the plate and ~80 Myr after acceleration of the plate (Farrington et al., 2010). Further work is required to determinate whether small-scale convection could explain part of the surface uplift observed in south Western Australia.

# 5. Conclusion

River profiles are sensitive to vertical motions and contain information pertaining to past uplift events (Whipple and Tucker, 1999). South Western Australian rivers consistently display a knickzone at ~200 ± 20 m, suggesting they recorded a common uplift history (Niemann et al., 2001). Applying an inverse algorithm (Pritchard et al., 2009), we show that these rivers have recorded ~200 m of surface uplift since the mid-Eocene. Predicted uplift rates increased from ~45-40 Ma onwards, and peaked at 22 m Myr$^{-1}$ during the Neogene. This surface uplift is consistent with the regional occurrence of elevated outcrops of shallow-marine Cenozoic sediments. We show that long-wavelength dynamic topography cannot solely account for this uplift history. However, the timing of changes in base level, defined as the difference between dynamic topography and eustasy, are consistent with the uplift history derived from longitudinal river profiles. Although secondary processes such as changes in far-field stresses and in climate may be at play, we propose that the post 45 Ma evolution of drainage basins in south Western Australia is mainly explained by eustasy and the migration of Australia away from a broad dynamic topography low related to the remnants of an ancient Pacific slab (Gurnis et al., 1998).

# Acknowledgements

NF was supported through an industry-grant from Statoil ASA. CH was funded by ARC Linkage LP0989312 supported by Shell and TOTAL. RDM was supported through ARC Laureate Fellowship FL0992245. This study benefit from discussions with Karol Czarnota, Guillaume Duclaux, and Jo Whittaker. All figures in this paper were generated using the open-source Generic Mapping Tools (Wessel et al., 2013). Some data was generated using the plate kinematic modelling software GPlates (http://www.gplates.org).

**Figure Captions**

**Figure 1:** Longitudinal river profiles (labelled 1- 18 and a – b) extracted from an SRTM 3s DEM (Rabus et al., 2003). Profiles are colour-coded by regions. South Western Australia (SWA) – orange, rivers that underwent documented drainage reversal – green, north Western Australia (NWA)– Blue, Northern Territories (NT) – Red, north Queensland (NQLD) – purple. Profile geometry varies across the continent. River profiles in Western Australia are underlain by the Darling Fault (white line) and the Perth Basin (cyan outline). SWA river profiles show major knickzones (Whipple and Tucker, 1999) that become less pronounced in NWA, from where profiles are concave. Inset: south Western Australian river profiles, major knickzones (red dots), 200 m contour elevation (white), and reactivated Late Neogene – Quaternary faults (yellow; Clark et al., 2012).

**Figure 2:** Parameter sensitivity study for $n$ = 1, knickpoint retreat rates $0.0125 < v_o < 60$ m $^{(1-m)}$ Myr$^{-1}$ and slope exponent $0.4 < m < 0.7$ for the Gardiner River. (a) Maximum magnitudes of uplift rate, U($\tau$), and (b) maximum characteristic uplift recording time, $\tau$. Both vary by one order of magnitude over the selected parameter range. Gold stars show the parameters used in this study, $m$ = 0.5 and $v_o$ = 5 m $^{(1-m)}$ Myr$^{-1}$ that results in maximum uplift rate < 10 m Myr$^{-1}$ and characteristic uplift recording time ∼120 Myr for the Gardiner river.

**Figure 3:** South Western Australian longitudinal river profiles and underlying bedrock lithology (Tyler and Hocking, 2008) divided based on erosional resistance (Roberts and White, 2010, Sklar and Dietrich, 2001). Note the absence of correlation between the location of major knickzones and change in underlying lithology, to the possible exception of the Phillips River (2).

**Figure 4:** Predicted uplift rate histories calculated using the method of Pritchard et al. (2009) for south Western Australian longitudinal river profiles. Uplift rates increase from ∼40 Ma, with maximum rates recorded during the Neogene. Exhumation deduced from apatite fission track analysis is shown in green

(green; Kohn et al., 2002), rates of dynamic topography change are shown in grey (Heine et al., 2010), and rate of base level change is shown in brown.

**Figure 5:** (a) Cumulative uplift normalised to present-day, for south Western Australian rivers. The present-day outcrop elevation of shallow-marine Eocene deposits (white box; Gammon et al., 2000) is used to calibrate $m$ and $v_o$. The present-day elevation (and estimated associated error) of shallow-marine Oligocene (Cope, 1975) and mid-Miocene (Lowry, 1970) is shown in cyan. (b) Location of shallow-marine Eocene deposits (gold star; Gammon et al., 2000) and 100 m and 200 m contours (grey).

**Figure. 6:** (a) Mid-Miocene marine deposits coloured by their present-day elevation, extent of offshore Mid-Miocene shallow-marine deposits (Nullarbor limestone and Plumridge limestone), south Western Australian river profiles (orange), major knickzones (red), onshore Eucla Basin (black) and Bremer (brown) basins, main outcrops of Pallinup Formation (gold stars) and 1888 – 2000 earthquakes of magnitude ≥ 5.5 (white circles; Dent, 2008) . (b) Latitudinal variations in elevations of Australian mid-Miocene limestones. Red stars represent median elevations for limestones north and south of 30$^0$S, respectively, consistent with a north-south continental scale tilt (Sandiford, 2007).

**Figure 7:** Dynamic topography predicted by a backward advection model at 0, 10, 20, 25, 30, and 40 Ma across the Australia continent (Heine et al., 2010). The Australian continent migrates across a broad-scale dynamic topography low, due to the sinking of the slab originating from the subduction zone that separated Eastern Gondwanaland from the paleo-Pacific Ocean during the Cretaceous (Gurnis et al., 1998). G. Evolution of long wavelength dynamic topography for south Western Australia. Hatched grey envelope represents range of values sampled at river mouths.  Evolution of long wavelength eustatic sea level (cyan; Haq and Al-Qahtani, 2005).

Fig. 1:

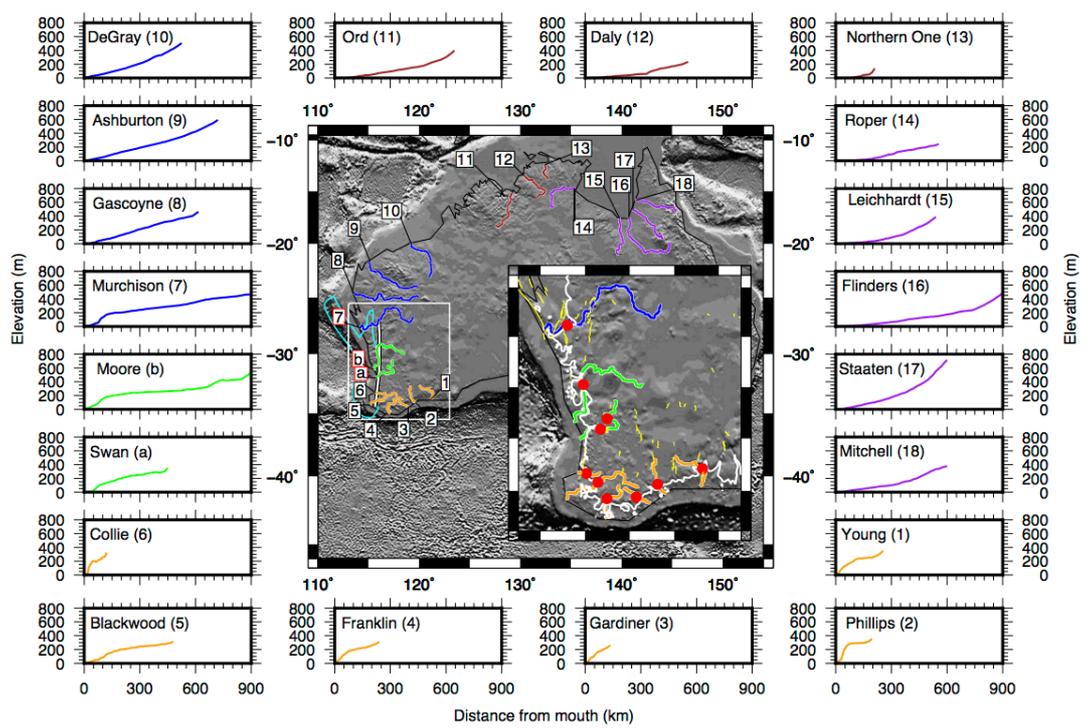

Fig. 2:

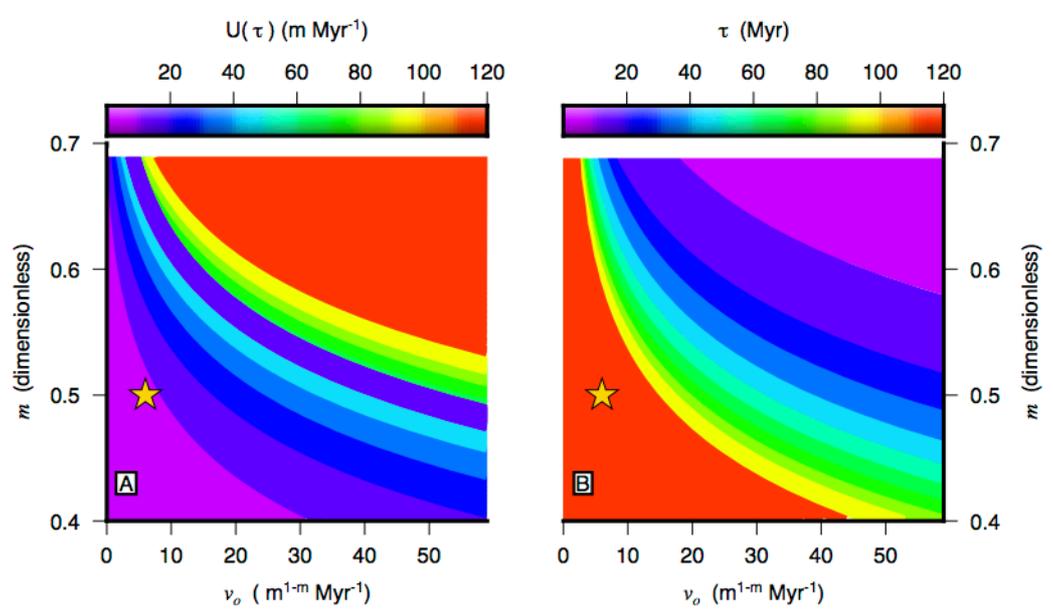

Fig. 3:

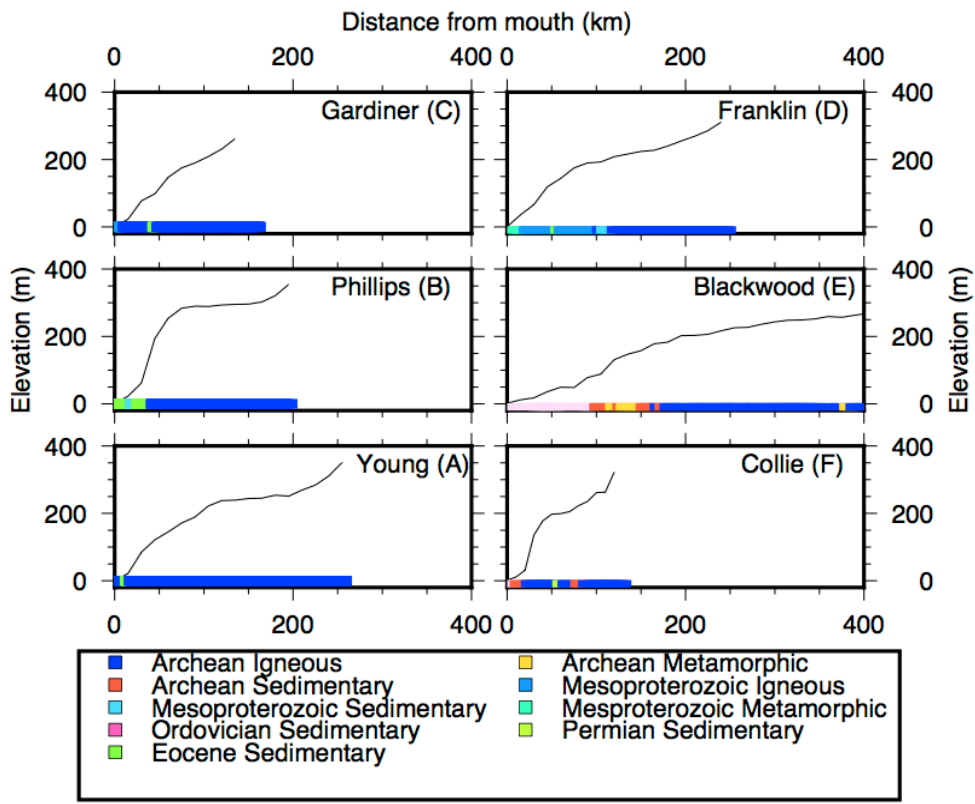

Fig. 4:

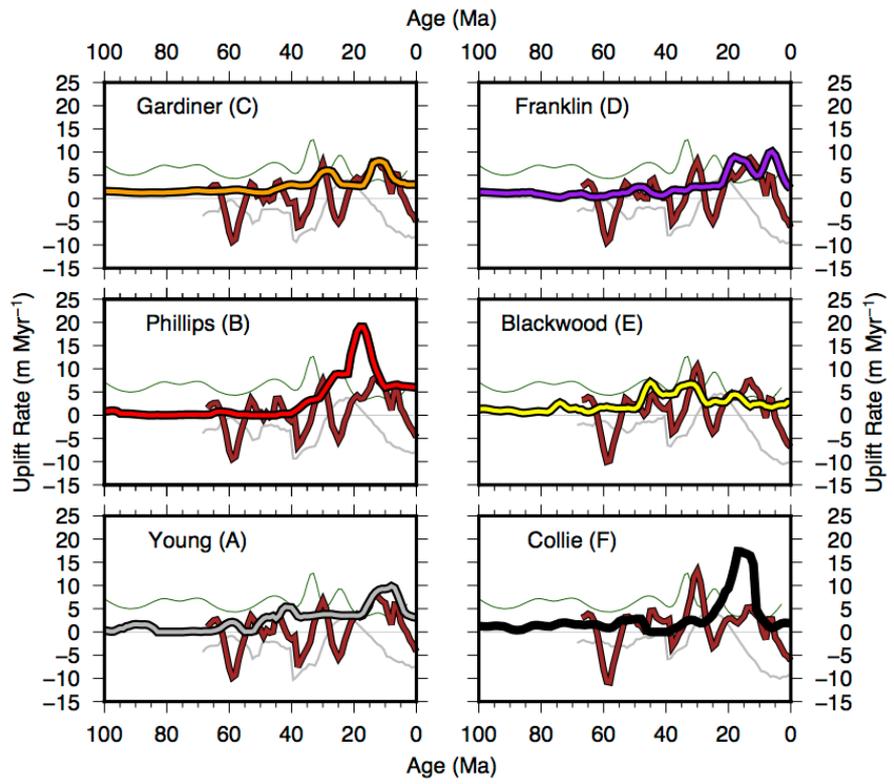

Fig. 5:

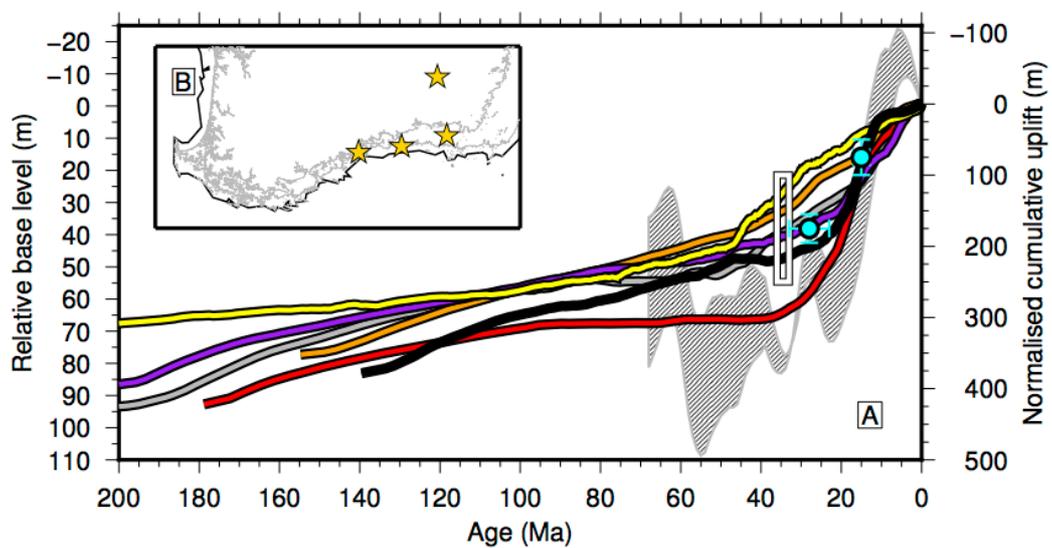

Fig. 6:

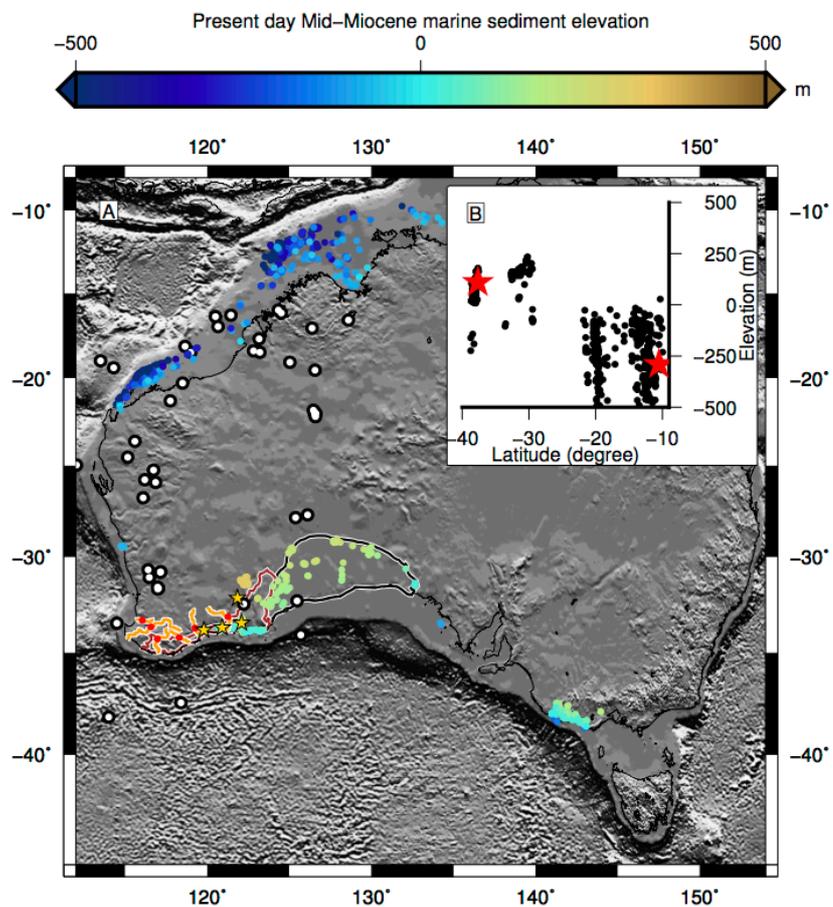

Fig. 7:

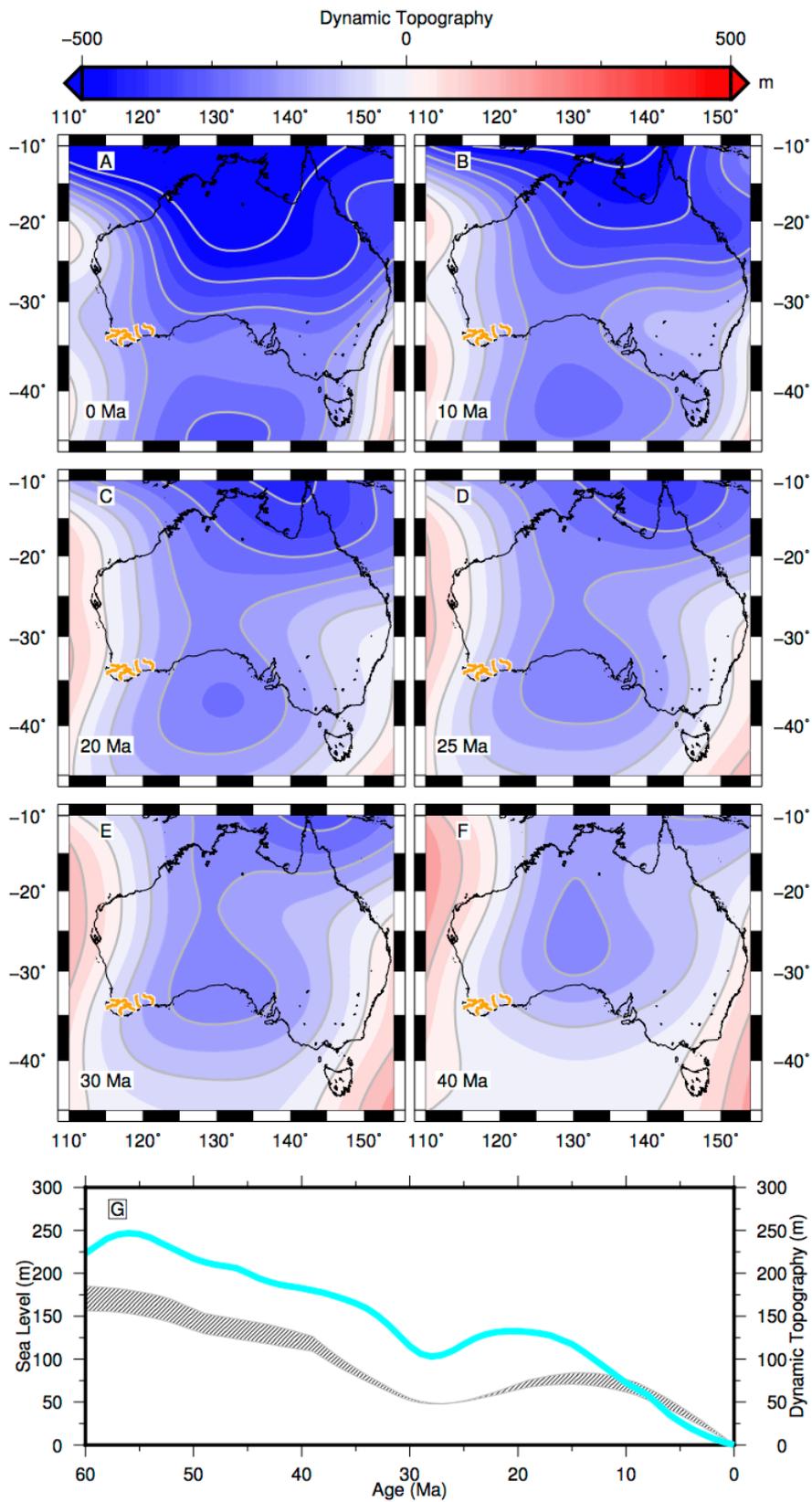